\documentclass[%
 reprint,
 amsmath,amssymb,
 aps,
]{revtex4-2}

\usepackage{graphicx}% Include figure files
\usepackage{dcolumn}% Align table columns on decimal point
\usepackage{bm}% bold math
\usepackage{hyperref}% add hypertext capabilities
\begin{document}

\preprint{APS/123-QED}

\title{Information extraction in photon-counting experiments}% Force line breaks with \\
%\thanks{A footnote to the article title}

\author{Timon Schapeler}\email{timon.schapeler@upb.de}
\author{Tim J. Bartley}
\affiliation{Mesoscopic Quantum Optics, Paderborn University, Warburger Straße 100, 33098 Paderborn, Germany}

\begin{abstract}
We demonstrate a comparison of different multiplexing architectures based on quantum detector tomography. Using the purity of their measurement outcomes, we gain insight about the photon-number resolving power of the devices. Further, we calculate the information each measurement outcome can extract out of a Hilbert space with given dimension. Our work confirms that more multiplexing outcomes enable higher photon-number resolving power; however, the splitting between those outcomes must be optimized as well.
\end{abstract}

\maketitle

%\tableofcontents

\section{Introduction}\label{sec:introduction}
The ability to count photons is crucial for quantum optical experiments and technologies, such as quantum metrology~\cite{you2021scalable}, quantum information~\cite{hadfield2009single}, and single-photon imaging~\cite{moreau2019imaging}. Many types of single-photon detectors have been demonstrated, all with certain advantages and disadvantages. The quality of these detectors can be quantified by many figures of merit, such as efficiency, dark counts or timing resolution~\cite{migdall2013single}. However, the capability of a detector to resolve the number of photons is a crucial quantifier in photon-counting experiments.

Resolving the photon-number can be fully achieved using for example superconducting transition-edge sensors (TESs)~\cite{cabrera1998detection}. For monochromatic light, a TES has full photon-number resolving capability, which means that a specific photon number maps to a specific output signal. Some information about photon number can be obtained using click-detectors such as superconducting nanowire single-photon detectors (SNSPDs) or single-photon avalanche diodes (SPADs) in combination with multiplexing schemes. However, multiplexing can only achieve quasi photon-number resolution, as some information is lost due to the nonzero probability of multiple photons causing the same outcome~\cite{kruse2017limits}.

Many different multiplexing architectures have been shown. Some of these schemes rely on equally splitting the incoming light onto the detectors, such as the conventional spatial~\cite{paul1996photon} and temporal~\cite{achilles2003fiber,fitch2003photon} multiplexing trees or detector arrays~\cite{dauler2007multi}. Others make use of logarithmic multiplexing such as the time-loop detectors~\cite{banaszek2003photon} or integrated in-line detector arrays~\cite{yu2018segmented}.

Multiplexed detectors can be parameterized by figures of merit such as efficiency, dark counts, and cross-talk~\cite{migdall2013single}. Although these quantities provide an intuition of the quality of the detectors, they do not quantify how these figures of merit combine to determine the utility of the devices for certain tasks. The photon-number resolving power of the device is one important example which is not directly described by these figures of merit. 

To overcome this limitation, van Enk introduced the concept of measurement outcome purity~\cite{vanEnk2017photodetector}, which can be directly linked to the ability of a detector architecture to resolve photon number. A related question arises: how much information can be extracted by a measurement outcome? In this paper, we experimentally investigate measurement outcome purity and information extraction for a variety of different detectors to quantify and compare their photon-number resolving power.

\begin{figure}[bht]
    \centering
    \includegraphics[width=0.99\linewidth]{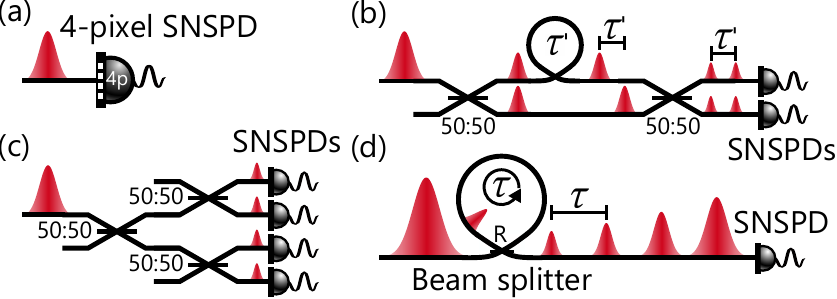}
    \caption{Schematic representations of (a) commercially available 4-pixel ($2\times2$) SNSPD array, (b) 4-bin  time-multiplexed detector (8 bins can be achieved with an additional fiber-loop of length $2\tau'$ and beam splitter), (c) 4-bin spatially-multiplexed detector, and (d) 10-bin time-multiplexed loop-detector with out-coupling $R$, bin separation $\tau$, and loop-efficiency $\eta_{\text{loop}}$, recently shown in~\cite{tiedau2019high}.}
    \label{fig:detectors}
\end{figure}

\section{Measurement outcome purity}\label{sec:background}
The specific detector architectures we consider are shown in Figs.~\ref{fig:detectors}(a)-\ref{fig:detectors}(d). In order to compare the different devices in a common framework, we use quantum detector tomography~\cite{lundeen2009tomography}, which yields a quantum mechanical description of a detector under test (DUT) in terms of its so called positive operator valued measures (POVMs). The set of POVMs $\{\pi_n\}$ fully describes the detector with different outcomes $n$. The operators are nonnegative $\pi_n>0$ with $\sum_n\pi_n=1$. 

An important property of a general detection scheme is that repeated measurements do not necessarily yield the same outcome, i.e. a measurement may not necessarily project onto pure states. Van Enk~\cite{vanEnk2017photodetector} showed that it is thus possible to define a measurement outcome purity of the POVM corresponding to outcome $n$, which is analogous to the purity of a quantum state, as
\begin{equation} \label{eq:Purity}
    \text{Pur}(\pi_n)=\frac{\text{Tr}((\pi_n)^2)}{(\text{Tr}(\pi_n))^2}\,.
\end{equation}
The purity is upper-bounded by unity, as a perfect measurement would be a one-to-one mapping of one input state to one outcome. The lower bound is given by the Hilbert space dimension $M$ as $\frac{1}{M}\leq\text{Pur}(\pi_n)\leq 1$. A non-pure POVM means that multiple orthogonal input states contribute to the same outcome with significant probabilities~\cite{vanEnk2017photodetector}. It is possible to estimate the number of orthogonal input states which contribute to outcome $n$ by the inverse of the purity $\text{Pur}(\pi_n)^{-1}$. This directly follows from the lower bound of the purity, where the minimal purity coincides with a contribution from every input state in the Hilbert space~\cite{nehra2020photon}.

\begin{figure}
    \centering
    \includegraphics[width=0.75\linewidth]{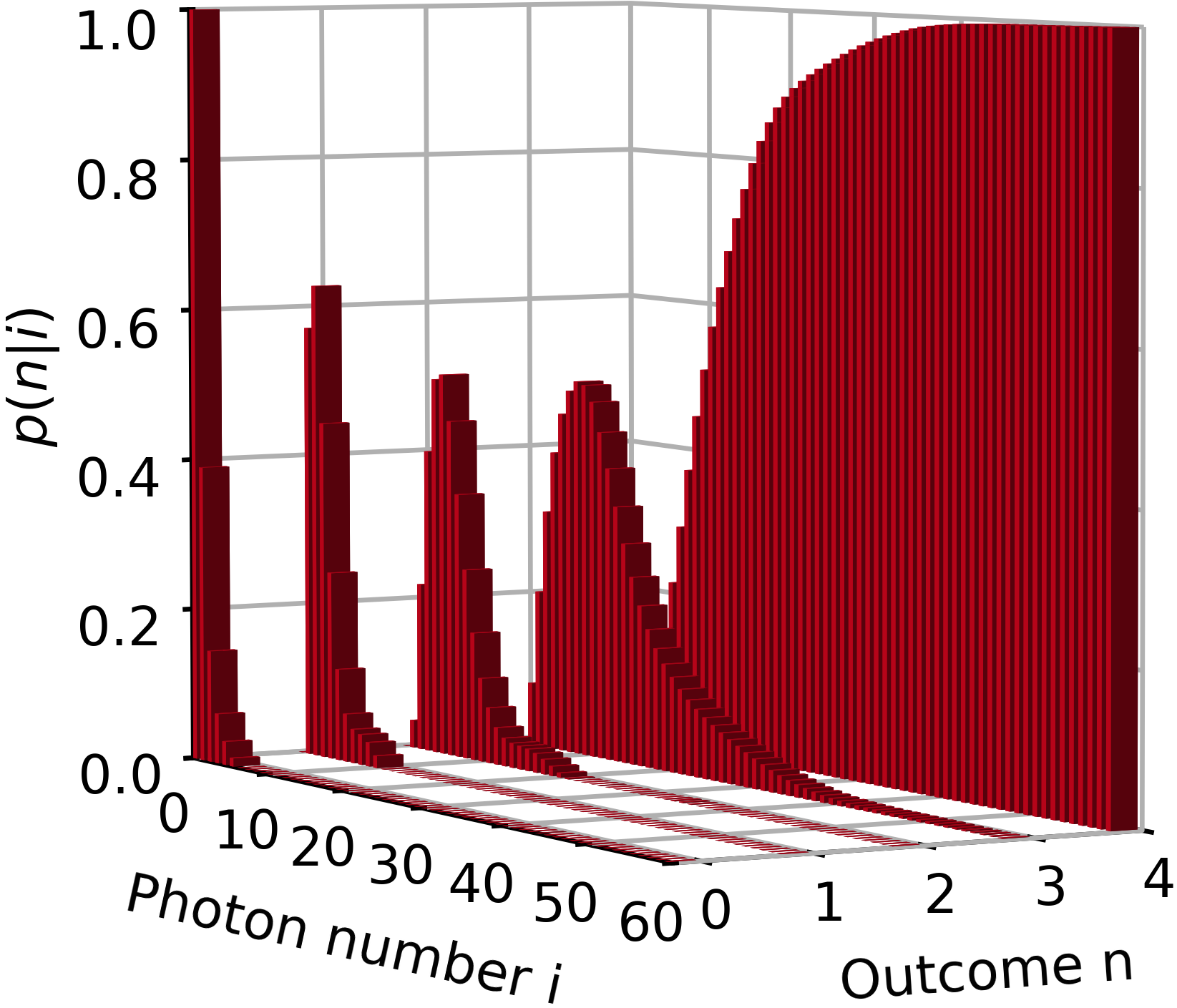}
    \caption{Diagonal elements of the reconstructed POVM operators in the photon-number basis for the 4-bin time-multiplexed detector, as an example to give a sense for the narrowness of outcomes and show saturation of the largest outcome.}
    \label{fig:POVMs}
\end{figure}

The investigated multiplexing schemes (shown in Figs.~\ref{fig:detectors}(a)-\ref{fig:detectors}(d)) have different dynamic ranges, which means they are sensitive to different numbers of photons. This translates to different Hilbert space dimensions $M$ in the POVM description. Above a certain photon number, the POVMs of a detector will not change. This can directly be seen in the outcome statistics, as with increasing mean photon numbers, the detector will only respond with the largest outcome. Therefore, the POVM corresponding to that outcome will saturate (occur with unit probability), while the other POVMs converge to zero probability. This can be seen in Fig.~\ref{fig:POVMs}, which shows the POVM elements for the 4-bin time-multiplexed detector (TMD) as an example.

\begin{figure}
    \centering
    \includegraphics[width=1\linewidth]{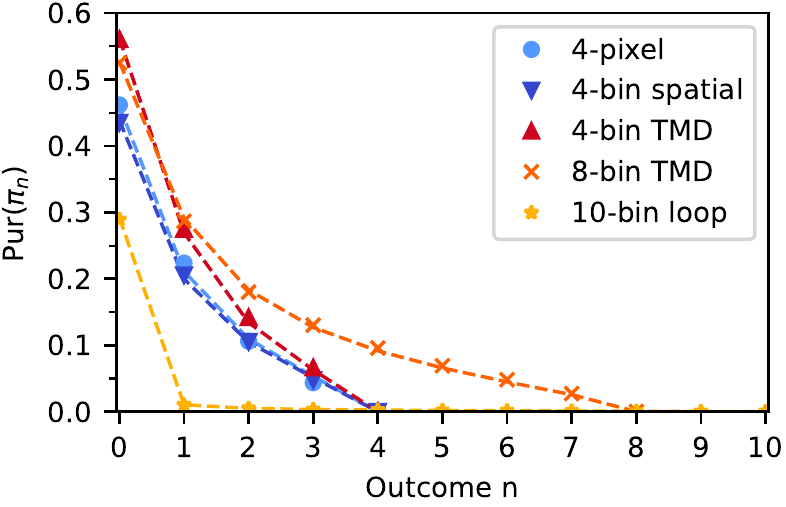}
    \caption{Purities $\text{Pur}(\pi_n)$ of the five multiplexed detectors per outcome $n$. Different colors and markers correspond to the different multiplexing schemes, as labeled in the legend. Dashed lines correspond to modeled data, see main text for details.}
    \label{fig:Purity}
\end{figure}

In order to compare the detector architectures, we use a fixed Hilbert space dimension of $M=5000$ without loss of generality, when assuming that the outcome statistics will not change after saturation of the largest outcome. We use Eq.~(\ref{eq:Purity}) to calculate the purities of the detectors, which are shown in Fig.~\ref{fig:Purity}. It can be seen that the detectors with four bins have comparable purities. This is expected, as the choice in the degree of freedom of multiplexing (in space or time) should not provide any significant advantages with regards to photon-number resolution. The small differences can be explained by slightly different performance metrics of the detectors (summarized in Table~\ref{tbl:results}), with the efficiency being the main contribution, as the purity suffers most for decreasing efficiencies~\cite{nehra2020photon}. Figure~\ref{fig:Purity} also shows that the outcome purities of the 8-bin TMD are higher compared to the 4-bin TMD. This makes sense, since fewer input states will contribute to each individual outcome, as more outcomes are available in the 8-bin TMD. Narrower POVM distributions imply higher purities.

We verified this behaviour by implementing a model of multiplexed click-detectors by Miatto et al.~\cite{miatto2018explicit}. We modeled the equal splitting devices (4-pixel, 4-bin and 8-bin TMD and 4-bin spatial) by solely their efficiencies (i.e. neglecting noise), which we extracted from the experimental data. The 10-bin loop detector model is based on previous work~\cite{schapeler2021quantum}.

The measurement outcome purities closely agree with the experimental data, as can be seen by the dashed lines with corresponding color in Fig.~\ref{fig:Purity}. However, more available outcomes does not directly imply higher outcome purities, as can be seen by the purities of the 10-bin time-multiplexed loop-detector in Fig.~\ref{fig:Purity}. This is due to the different multiplexing scheme. The device is based on logarithmic multiplexing, while the other architectures rely on equal splitting. The 10-bin loop-detector responds logarithmically to the number of photons, which results in broad POVM distributions (many orthogonal input states contribute to a given outcome) and thus in poor purities. 

\section{Information extraction}\label{sec:InfExt}
The purity of a photon counting detection outcome can be intuitively regarded as the photon-number resolving power of the device, or the ability of the detector to distinguish between different photon-number states. A related concept is information extraction, namely how much information about the photon number can be obtained given a specific outcome. Here, van Enk~\cite{vanEnk2017photodetector} introduced the entropy
\begin{equation} \label{eq:H_missing}
    H^{(n)} = -\sum_i p(i|n)\log_2(p(i|n)) \,,
\end{equation}
which quantifies the information (in bits) about the photon number $i$ that is still missing after having obtained the outcome $n$. This quantity uses the conditional probability $p(i|n)$, which describes the probability of having a photon number equal to $i$ given a measurement outcome $n$.

We are able to obtain the probability $p(i|n)$, utilizing the POVMs of the detectors. A single element of the POVM $\pi_n$ is the conditional probability $p(n|i)$ of $n$ clicks occurring given $i$ incident photons. Using Bayes' theorem
\begin{equation} 
    p(i|n) = \frac{p(n|i)p(i)}{p(n)} = \frac{p(n|i)\sum_{i,n} p(n|i)}{M\sum_i p(n|i)} \,,
\end{equation}
it is possible to obtain the probability $p(i|n)$ directly from the POVMs. Here we assume a flat prior probability $p(i)=\frac{1}{M}$, meaning that all input states are equally likely. 

From Eq.~(\ref{eq:H_missing}) we can calculate the information that can be extracted by a certain outcome $n$ about the photon number $i$ as
\begin{equation} \label{eq:H_extr}
    H_{\text{extr}}^{(n)} = H_{\text{total}}(M) - H^{(n)} \,,
\end{equation}
where $H_{\text{total}}(M) = - \log_2\left(\frac{1}{M}\right) = \log_2(M)$ is the total information contained in the Hilbert space of dimension $M$. 

\begin{figure}
    \centering
    \includegraphics[width=1\linewidth]{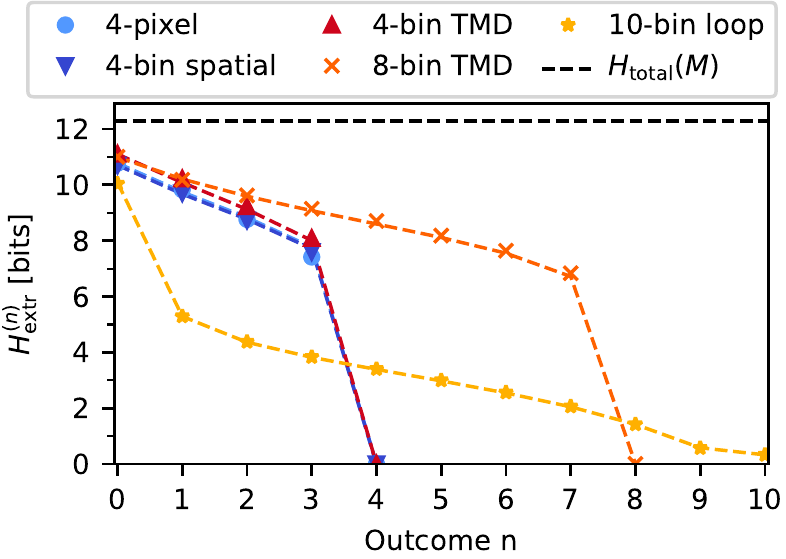}
    \caption{Extracted information $H_{\text{extr}}^{(n)}$ in bits of the five multiplexed detectors per outcome $n$. The total available information in the Hilbert space of dimension $M=5000$ is shown by the black dashed line. Different colors and markers correspond to the different multiplexing schemes, as labeled in the legend. Dashed lines correspond to modeled data, see main text for details.}
    \label{fig:Info}
\end{figure}

Using Eq.~(\ref{eq:H_extr}) we calculate the amount of information that can be extracted out of the Hilbert space of size $M=5000$. The total information available calculates to $H_{\text{total}}(5000)=12.3$ bits, which is shown by the black dashed line in Fig.~\ref{fig:Info}. Having similar performance metrics and outcome purities, the detectors with four bins extract the same amount of information out of the Hilbert space. Figure~\ref{fig:Info} also shows that the outcomes of the 8-bin TMD extract the most information. This directly follows from the argument that more available outcomes lead to narrower POVMs and make each outcome more specific. If a narrow outcome is observed, more knowledge about the photon number can be extracted, as less photon numbers (or orthogonal input states) contribute to this outcome. This also explains the sudden drop in the extracted information for the largest outcome of the detectors based on equal splitting. These detectors have a low dynamic range (roughly $M\approx100$) compared to the maximum Hilbert space of size $M=5000$. This means that obtaining the largest outcome reveals very little about the possible input photon number, due to the saturation of the outcome (compare Fig.~\ref{fig:POVMs}). 

The outcomes of the 10-bin loop-detector, based on logarithmic multiplexing, span the Hilbert space with broad POVM distributions. This leads to less information extracted per outcome. However, even the largest outcome is able to extract some information out of the total available Hilbert space, even for large input states with hundreds to thousands of photons.
From this, it is clear that solely increasing the number of multiplexing outcomes will not yield better photon-number resolution. Optimizing the photon-number resolving power of a device requires optimizing the splitting between multiplexed detectors~\cite{nehra2020photon}.

\section{Experimental methods}\label{sec:methods}
\begin{figure}[ht]
    \centering
    \includegraphics[width=1\linewidth]{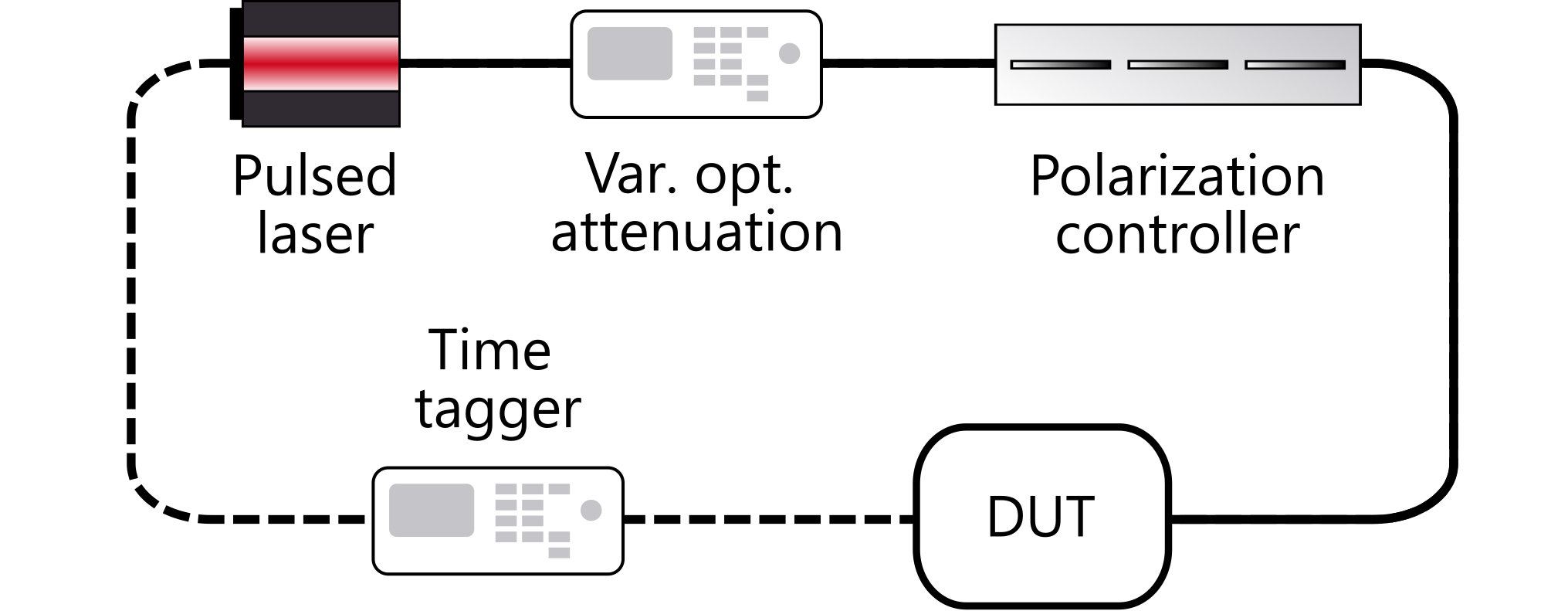}
    \caption{Experimental setup for the detector tomography analysis. A 1556~nm pulsed laser produces coherent states, which can be attenuated by variable optical attenuators. The polarization can be controlled to optimize detection efficiencies for the polarization dependent SNSPDs, which are connected next. A time tagger records the electrical responses from the detectors. Solid and dashed lines represent optical and electrical connections respectively.}
    \label{fig:setup}
\end{figure}
Calculating the purity and information extraction relies on a tomographic reconstruction of each detector. We summarize the methods here; more detail can be found in Refs.~\cite{lundeen2009tomography,feito2009measuring,schapeler2020quantum,schapeler2021quantum}.

In order to  reconstruct the POVMs, the detector under test needs to be subjected to a set of input states spanning the Hilbert space of the given device. Coherent states are an ideal choice, as they form a tomographically complete set~\cite{lundeen2009tomography}. Here we compare four different multiplexing schemes (five detectors) in total: a commercially available four-pixel SNSPD array, a 4-bin (and 8-bin) time-multiplexed detector (TMD), a 4-bin spatially-multiplexed detector, and finally a 10-bin time-multiplexed loop-detector, which relies on a logarithmic time-multiplexing architecture enabling a high dynamic-range over 120~dB~\cite{tiedau2019high} (all shown in Figs.~\ref{fig:detectors}(a)-\ref{fig:detectors}(d)). The experimental setup, shown schematically in Fig.~\ref{fig:setup}, begins with a pulsed laser producing coherent states with a wavelength of 1556~nm. The mean photon number can be controlled by variable optical attenuators and the polarization can be set to optimize the detection efficiency of the polarization dependent SNSPDs. Afterwards, the different multiplexing schemes are connected and the outcome statistics are recorded by a time tagger in a few-nanosecond wide coincidence window. The mean photon numbers of the coherent input states are chosen to scale quadratically to efficiently span the Hilbert space of the different detectors. Different amounts of input states are necessary, as the Hilbert space dimension depends on the detector under test.

After having obtained the outcome statistics, typically represented by a matrix $\mathbf{P}$, and knowing the input states, typically represented by a matrix $\mathbf{F}$, it is possible to reconstruct the POVM matrix $\mathbf{\Pi}$ of the different multiplexed detectors. This is done using the Born rule, which allows to formulate the matrix equation $\mathbf{P}=\mathbf{F}\mathbf{\Pi}$ and a subsequent matrix inversion routine~\cite{schapeler2020cvxpycode,cvxpy,cvxpy_rewriting}.

\begin{table}[t]
 \centering 
 \caption{Figures of merit (efficiency $\eta$, dark-count probability $p_{\mathrm{dark}}$ and cross-talk probability $p_{\mathrm{xtalk}}$) of the five different multiplexed detectors, obtained from the POVMs as in Ref.~\cite{schapeler2020quantum}. The errors are based on assuming $5\%$ uncertainty in the amplitudes of the coherent states.}
\begin{tabular}{l|l|l|l}
    \hline
    \hline
                    & $\eta$ [\%]   & $p_{\mathrm{dark}}$ [\%]        & $p_{\mathrm{xtalk}}$ [\%]\\
    \hline
    4-pixel         & $63\pm 4$   & $(5.9\pm 1.6)\times 10^{-4}$    & $14\pm 1$\\
    4-bin spatial   & $61\pm 4$   & $(5.4\pm 1.8)\times 10^{-4}$    & $3.6\pm 0.1$\\
    4-bin TMD       & $72\pm 4$   & $(1.6\pm 0.1)\times 10^{-4}$    & $(0 + 2)\times 10^{-4}$\\
    8-bin TMD       & $69\pm 4$   & $<10^{-5}$                      & $(4.3\pm 1.1)\times 10^{-5}$\\
    10-bin loop     & $44\pm 3$   & $(0.0 + 4.1)\times 10^{-5}$     & $(4.6\pm 2.8)\times 10^{-4}$\\
    \hline
    \hline
\end{tabular} \label{tbl:results}
\end{table}

\section{Conclusion}\label{sec:conclusion}
We have shown that quantum detector tomography is the basis to enable a comparison of different multiplexing schemes. The outcome purity, which can be calculated directly from the POVM elements of given detection outcomes, indicates how pure a certain outcome is. %The fewer orthogonal input states contribute to a given outcome, the purer this outcome is.
Quantum detector tomography enables a mapping to photon number states, thus the purity gives insight about the photon-number resolving power of the devices. Narrow POVM distributions are purer, in the sense that fewer photon numbers contribute to the given outcome. This directly translates into the amount of information the outcome can extract out of the total available Hilbert space. More information can be extracted by narrow POVM distributions. Outcomes of multiplexing architectures based on equal splitting, such as standard spatial or temporal multiplexing, show significantly better purities and can extract larger amounts of information, compared to the logarithmic multiplexing scheme of the 10-bin loop-detector. Nevertheless, logarithmic multiplexing enables a high dynamic range, which gives information about large photon numbers, where the equal splitting schemes are already saturated.
Our analysis further confirms that more multiplexing outcomes enable higher photon-number resolving power~\cite{sperling2012true,kruse2017limits,jonsson2019evaluating}, but also that the splitting between those outcomes must be optimized.

\section*{Acknowledgements}
This work was supported by the Bundesministerium für Bildung und Forschung (13N14911).

\bibliography{sources}% Produces the bibliography via BibTeX.

\begin{thebibliography}{25}%
\makeatletter
\providecommand \@ifxundefined [1]{%
 \@ifx{#1\undefined}
}%
\providecommand \@ifnum [1]{%
 \ifnum #1\expandafter \@firstoftwo
 \else \expandafter \@secondoftwo
 \fi
}%
\providecommand \@ifx [1]{%
 \ifx #1\expandafter \@firstoftwo
 \else \expandafter \@secondoftwo
 \fi
}%
\providecommand \natexlab [1]{#1}%
\providecommand \enquote  [1]{``#1''}%
\providecommand \bibnamefont  [1]{#1}%
\providecommand \bibfnamefont [1]{#1}%
\providecommand \citenamefont [1]{#1}%
\providecommand \href@noop [0]{\@secondoftwo}%
\providecommand \href [0]{\begingroup \@sanitize@url \@href}%
\providecommand \@href[1]{\@@startlink{#1}\@@href}%
\providecommand \@@href[1]{\endgroup#1\@@endlink}%
\providecommand \@sanitize@url [0]{\catcode `\\12\catcode `\$12\catcode
  `\&12\catcode `\#12\catcode `\^12\catcode `\_12\catcode `\%12\relax}%
\providecommand \@@startlink[1]{}%
\providecommand \@@endlink[0]{}%
\providecommand \url  [0]{\begingroup\@sanitize@url \@url }%
\providecommand \@url [1]{\endgroup\@href {#1}{\urlprefix }}%
\providecommand \urlprefix  [0]{URL }%
\providecommand \Eprint [0]{\href }%
\providecommand \doibase [0]{https://doi.org/}%
\providecommand \selectlanguage [0]{\@gobble}%
\providecommand \bibinfo  [0]{\@secondoftwo}%
\providecommand \bibfield  [0]{\@secondoftwo}%
\providecommand \translation [1]{[#1]}%
\providecommand \BibitemOpen [0]{}%
\providecommand \bibitemStop [0]{}%
\providecommand \bibitemNoStop [0]{.\EOS\space}%
\providecommand \EOS [0]{\spacefactor3000\relax}%
\providecommand \BibitemShut  [1]{\csname bibitem#1\endcsname}%
\let\auto@bib@innerbib\@empty
%</preamble>
\bibitem [{\citenamefont {You}\ \emph {et~al.}(2021)\citenamefont {You},
  \citenamefont {Hong}, \citenamefont {Bierhorst}, \citenamefont {Lita},
  \citenamefont {Glancy}, \citenamefont {Kolthammer}, \citenamefont {Knill},
  \citenamefont {Nam}, \citenamefont {Mirin}, \citenamefont
  {Maga{\~{n}}a-Loaiza},\ and\ \citenamefont {Gerrits}}]{you2021scalable}%
  \BibitemOpen
  \bibfield  {author} {\bibinfo {author} {\bibfnamefont {C.}~\bibnamefont
  {You}}, \bibinfo {author} {\bibfnamefont {M.}~\bibnamefont {Hong}}, \bibinfo
  {author} {\bibfnamefont {P.}~\bibnamefont {Bierhorst}}, \bibinfo {author}
  {\bibfnamefont {A.~E.}\ \bibnamefont {Lita}}, \bibinfo {author}
  {\bibfnamefont {S.}~\bibnamefont {Glancy}}, \bibinfo {author} {\bibfnamefont
  {S.}~\bibnamefont {Kolthammer}}, \bibinfo {author} {\bibfnamefont
  {E.}~\bibnamefont {Knill}}, \bibinfo {author} {\bibfnamefont {S.~W.}\
  \bibnamefont {Nam}}, \bibinfo {author} {\bibfnamefont {R.~P.}\ \bibnamefont
  {Mirin}}, \bibinfo {author} {\bibfnamefont {O.~S.}\ \bibnamefont
  {Maga{\~{n}}a-Loaiza}},\ and\ \bibinfo {author} {\bibfnamefont
  {T.}~\bibnamefont {Gerrits}},\ }\bibfield  {title} {\bibinfo {title}
  {{Scalable multiphoton quantum metrology with neither pre- nor post-selected
  measurements}},\ }\href {https://doi.org/10.1063/5.0063294} {\bibfield
  {journal} {\bibinfo  {journal} {Applied Physics Reviews}\ }\textbf {\bibinfo
  {volume} {8}},\ \bibinfo {pages} {41406} (\bibinfo {year}
  {2021})}\BibitemShut {NoStop}%
\bibitem [{\citenamefont {Hadfield}(2009)}]{hadfield2009single}%
  \BibitemOpen
  \bibfield  {author} {\bibinfo {author} {\bibfnamefont {R.~H.}\ \bibnamefont
  {Hadfield}},\ }\bibfield  {title} {\bibinfo {title} {{Single-photon detectors
  for optical quantum information applications}},\ }\href
  {https://doi.org/10.1038/nphoton.2009.230} {\bibfield  {journal} {\bibinfo
  {journal} {Nature Photonics}\ }\textbf {\bibinfo {volume} {3}},\ \bibinfo
  {pages} {696} (\bibinfo {year} {2009})}\BibitemShut {NoStop}%
\bibitem [{\citenamefont {Moreau}\ \emph {et~al.}(2019)\citenamefont {Moreau},
  \citenamefont {Toninelli}, \citenamefont {Gregory},\ and\ \citenamefont
  {Padgett}}]{moreau2019imaging}%
  \BibitemOpen
  \bibfield  {author} {\bibinfo {author} {\bibfnamefont {P.-A.}\ \bibnamefont
  {Moreau}}, \bibinfo {author} {\bibfnamefont {E.}~\bibnamefont {Toninelli}},
  \bibinfo {author} {\bibfnamefont {T.}~\bibnamefont {Gregory}},\ and\ \bibinfo
  {author} {\bibfnamefont {M.~J.}\ \bibnamefont {Padgett}},\ }\bibfield
  {title} {\bibinfo {title} {{Imaging with quantum states of light}},\ }\href
  {https://doi.org/10.1038/s42254-019-0056-0} {\bibfield  {journal} {\bibinfo
  {journal} {Nature Reviews Physics}\ }\textbf {\bibinfo {volume} {1}},\
  \bibinfo {pages} {367} (\bibinfo {year} {2019})}\BibitemShut {NoStop}%
\bibitem [{\citenamefont {Migdall}\ \emph {et~al.}(2013)\citenamefont
  {Migdall}, \citenamefont {Polyakov}, \citenamefont {Fan},\ and\ \citenamefont
  {Bienfang}}]{migdall2013single}%
  \BibitemOpen
  \bibfield  {author} {\bibinfo {author} {\bibfnamefont {A.}~\bibnamefont
  {Migdall}}, \bibinfo {author} {\bibfnamefont {S.~V.}\ \bibnamefont
  {Polyakov}}, \bibinfo {author} {\bibfnamefont {J.}~\bibnamefont {Fan}},\ and\
  \bibinfo {author} {\bibfnamefont {J.~C.}\ \bibnamefont {Bienfang}},\ }\href
  {https://books.google.de/books?id=gERMqvh2OjAC} {\emph {\bibinfo {title}
  {Single-Photon Generation and Detection: Physics and Applications}}}\
  (\bibinfo  {publisher} {Academic Press},\ \bibinfo {year} {2013})\BibitemShut
  {NoStop}%
\bibitem [{\citenamefont {Cabrera}\ \emph {et~al.}(1998)\citenamefont
  {Cabrera}, \citenamefont {Clarke}, \citenamefont {Colling}, \citenamefont
  {Miller}, \citenamefont {Nam},\ and\ \citenamefont
  {Romani}}]{cabrera1998detection}%
  \BibitemOpen
  \bibfield  {author} {\bibinfo {author} {\bibfnamefont {B.}~\bibnamefont
  {Cabrera}}, \bibinfo {author} {\bibfnamefont {R.~M.}\ \bibnamefont {Clarke}},
  \bibinfo {author} {\bibfnamefont {P.}~\bibnamefont {Colling}}, \bibinfo
  {author} {\bibfnamefont {A.~J.}\ \bibnamefont {Miller}}, \bibinfo {author}
  {\bibfnamefont {S.}~\bibnamefont {Nam}},\ and\ \bibinfo {author}
  {\bibfnamefont {R.~W.}\ \bibnamefont {Romani}},\ }\bibfield  {title}
  {\bibinfo {title} {{Detection of single infrared, optical, and ultraviolet
  photons using superconducting transition edge sensors}},\ }\href
  {https://doi.org/10.1063/1.121984} {\bibfield  {journal} {\bibinfo  {journal}
  {Applied Physics Letters}\ }\textbf {\bibinfo {volume} {73}},\ \bibinfo
  {pages} {735} (\bibinfo {year} {1998})}\BibitemShut {NoStop}%
\bibitem [{\citenamefont {Kruse}\ \emph {et~al.}(2017)\citenamefont {Kruse},
  \citenamefont {Tiedau}, \citenamefont {Bartley}, \citenamefont {Barkhofen},\
  and\ \citenamefont {Silberhorn}}]{kruse2017limits}%
  \BibitemOpen
  \bibfield  {author} {\bibinfo {author} {\bibfnamefont {R.}~\bibnamefont
  {Kruse}}, \bibinfo {author} {\bibfnamefont {J.}~\bibnamefont {Tiedau}},
  \bibinfo {author} {\bibfnamefont {T.~J.}\ \bibnamefont {Bartley}}, \bibinfo
  {author} {\bibfnamefont {S.}~\bibnamefont {Barkhofen}},\ and\ \bibinfo
  {author} {\bibfnamefont {C.}~\bibnamefont {Silberhorn}},\ }\bibfield  {title}
  {\bibinfo {title} {{Limits of the time-multiplexed photon-counting method}},\
  }\href {https://doi.org/10.1103/PhysRevA.95.023815} {\bibfield  {journal}
  {\bibinfo  {journal} {Physical Review A}\ }\textbf {\bibinfo {volume} {95}},\
  \bibinfo {pages} {23815} (\bibinfo {year} {2017})}\BibitemShut {NoStop}%
\bibitem [{\citenamefont {Paul}\ \emph {et~al.}(1996)\citenamefont {Paul},
  \citenamefont {T{\"{o}}rm{\"{a}}}, \citenamefont {Kiss},\ and\ \citenamefont
  {Jex}}]{paul1996photon}%
  \BibitemOpen
  \bibfield  {author} {\bibinfo {author} {\bibfnamefont {H.}~\bibnamefont
  {Paul}}, \bibinfo {author} {\bibfnamefont {P.}~\bibnamefont
  {T{\"{o}}rm{\"{a}}}}, \bibinfo {author} {\bibfnamefont {T.}~\bibnamefont
  {Kiss}},\ and\ \bibinfo {author} {\bibfnamefont {I.}~\bibnamefont {Jex}},\
  }\bibfield  {title} {\bibinfo {title} {{Photon Chopping: New Way to Measure
  the Quantum State of Light}},\ }\href
  {https://doi.org/10.1103/PhysRevLett.76.2464} {\bibfield  {journal} {\bibinfo
   {journal} {Physical Review Letters}\ }\textbf {\bibinfo {volume} {76}},\
  \bibinfo {pages} {2464} (\bibinfo {year} {1996})}\BibitemShut {NoStop}%
\bibitem [{\citenamefont {Achilles}\ \emph {et~al.}(2003)\citenamefont
  {Achilles}, \citenamefont {Silberhorn}, \citenamefont {{\'{S}}liwa},
  \citenamefont {Banaszek},\ and\ \citenamefont
  {Walmsley}}]{achilles2003fiber}%
  \BibitemOpen
  \bibfield  {author} {\bibinfo {author} {\bibfnamefont {D.}~\bibnamefont
  {Achilles}}, \bibinfo {author} {\bibfnamefont {C.}~\bibnamefont
  {Silberhorn}}, \bibinfo {author} {\bibfnamefont {C.}~\bibnamefont
  {{\'{S}}liwa}}, \bibinfo {author} {\bibfnamefont {K.}~\bibnamefont
  {Banaszek}},\ and\ \bibinfo {author} {\bibfnamefont {I.~A.}\ \bibnamefont
  {Walmsley}},\ }\bibfield  {title} {\bibinfo {title} {{Fiber-assisted
  detection with photon number resolution}},\ }\href
  {https://doi.org/10.1364/OL.28.002387} {\bibfield  {journal} {\bibinfo
  {journal} {Optics Letters}\ }\textbf {\bibinfo {volume} {28}},\ \bibinfo
  {pages} {2387} (\bibinfo {year} {2003})}\BibitemShut {NoStop}%
\bibitem [{\citenamefont {Fitch}\ \emph {et~al.}(2003)\citenamefont {Fitch},
  \citenamefont {Jacobs}, \citenamefont {Pittman},\ and\ \citenamefont
  {Franson}}]{fitch2003photon}%
  \BibitemOpen
  \bibfield  {author} {\bibinfo {author} {\bibfnamefont {M.~J.}\ \bibnamefont
  {Fitch}}, \bibinfo {author} {\bibfnamefont {B.~C.}\ \bibnamefont {Jacobs}},
  \bibinfo {author} {\bibfnamefont {T.~B.}\ \bibnamefont {Pittman}},\ and\
  \bibinfo {author} {\bibfnamefont {J.~D.}\ \bibnamefont {Franson}},\
  }\bibfield  {title} {\bibinfo {title} {{Photon-number resolution using
  time-multiplexed single-photon detectors}},\ }\href
  {https://doi.org/10.1103/PhysRevA.68.043814} {\bibfield  {journal} {\bibinfo
  {journal} {Physical Review A}\ }\textbf {\bibinfo {volume} {68}},\ \bibinfo
  {pages} {43814} (\bibinfo {year} {2003})}\BibitemShut {NoStop}%
\bibitem [{\citenamefont {Dauler}\ \emph {et~al.}(2007)\citenamefont {Dauler},
  \citenamefont {Robinson}, \citenamefont {Kerman}, \citenamefont {Yang},
  \citenamefont {Rosfjord}, \citenamefont {Anant}, \citenamefont {Voronov},
  \citenamefont {Gol'tsman},\ and\ \citenamefont {Berggren}}]{dauler2007multi}%
  \BibitemOpen
  \bibfield  {author} {\bibinfo {author} {\bibfnamefont {E.~A.}\ \bibnamefont
  {Dauler}}, \bibinfo {author} {\bibfnamefont {B.~S.}\ \bibnamefont
  {Robinson}}, \bibinfo {author} {\bibfnamefont {A.~J.}\ \bibnamefont
  {Kerman}}, \bibinfo {author} {\bibfnamefont {J.~K.~W.}\ \bibnamefont {Yang}},
  \bibinfo {author} {\bibfnamefont {K.~M.}\ \bibnamefont {Rosfjord}}, \bibinfo
  {author} {\bibfnamefont {V.}~\bibnamefont {Anant}}, \bibinfo {author}
  {\bibfnamefont {B.}~\bibnamefont {Voronov}}, \bibinfo {author} {\bibfnamefont
  {G.}~\bibnamefont {Gol'tsman}},\ and\ \bibinfo {author} {\bibfnamefont
  {K.~K.}\ \bibnamefont {Berggren}},\ }\bibfield  {title} {\bibinfo {title}
  {{Multi-Element Superconducting Nanowire Single-Photon Detector}},\ }\href
  {https://doi.org/10.1109/TASC.2007.897372} {\bibfield  {journal} {\bibinfo
  {journal} {IEEE Transactions on Applied Superconductivity}\ }\textbf
  {\bibinfo {volume} {17}},\ \bibinfo {pages} {279} (\bibinfo {year}
  {2007})}\BibitemShut {NoStop}%
\bibitem [{\citenamefont {Banaszek}\ and\ \citenamefont
  {Walmsley}(2003)}]{banaszek2003photon}%
  \BibitemOpen
  \bibfield  {author} {\bibinfo {author} {\bibfnamefont {K.}~\bibnamefont
  {Banaszek}}\ and\ \bibinfo {author} {\bibfnamefont {I.~A.}\ \bibnamefont
  {Walmsley}},\ }\bibfield  {title} {\bibinfo {title} {{Photon counting with a
  loop detector}},\ }\href {https://doi.org/10.1364/OL.28.000052} {\bibfield
  {journal} {\bibinfo  {journal} {Optics Letters}\ }\textbf {\bibinfo {volume}
  {28}},\ \bibinfo {pages} {52} (\bibinfo {year} {2003})}\BibitemShut {NoStop}%
\bibitem [{\citenamefont {Yu}\ \emph {et~al.}(2018)\citenamefont {Yu},
  \citenamefont {Sun}, \citenamefont {Li},\ and\ \citenamefont
  {Beling}}]{yu2018segmented}%
  \BibitemOpen
  \bibfield  {author} {\bibinfo {author} {\bibfnamefont {Q.}~\bibnamefont
  {Yu}}, \bibinfo {author} {\bibfnamefont {K.}~\bibnamefont {Sun}}, \bibinfo
  {author} {\bibfnamefont {Q.}~\bibnamefont {Li}},\ and\ \bibinfo {author}
  {\bibfnamefont {A.}~\bibnamefont {Beling}},\ }\bibfield  {title} {\bibinfo
  {title} {{Segmented waveguide photodetector with 90{\%} quantum
  efficiency}},\ }\href {https://doi.org/10.1364/OE.26.012499} {\bibfield
  {journal} {\bibinfo  {journal} {Optics Express}\ }\textbf {\bibinfo {volume}
  {26}},\ \bibinfo {pages} {12499} (\bibinfo {year} {2018})}\BibitemShut
  {NoStop}%
\bibitem [{\citenamefont {van Enk}(2017)}]{vanEnk2017photodetector}%
  \BibitemOpen
  \bibfield  {author} {\bibinfo {author} {\bibfnamefont {S.~J.}\ \bibnamefont
  {van Enk}},\ }\bibfield  {title} {\bibinfo {title} {Photodetector figures of
  merit in terms of {POVMs}},\ }\href
  {https://doi.org/10.1088/2399-6528/aa90ce} {\bibfield  {journal} {\bibinfo
  {journal} {Journal of Physics Communications}\ }\textbf {\bibinfo {volume}
  {1}},\ \bibinfo {pages} {45001} (\bibinfo {year} {2017})}\BibitemShut
  {NoStop}%
\bibitem [{\citenamefont {Tiedau}\ \emph {et~al.}(2019)\citenamefont {Tiedau},
  \citenamefont {Meyer-Scott}, \citenamefont {Nitsche}, \citenamefont
  {Barkhofen}, \citenamefont {Bartley},\ and\ \citenamefont
  {Silberhorn}}]{tiedau2019high}%
  \BibitemOpen
  \bibfield  {author} {\bibinfo {author} {\bibfnamefont {J.}~\bibnamefont
  {Tiedau}}, \bibinfo {author} {\bibfnamefont {E.}~\bibnamefont {Meyer-Scott}},
  \bibinfo {author} {\bibfnamefont {T.}~\bibnamefont {Nitsche}}, \bibinfo
  {author} {\bibfnamefont {S.}~\bibnamefont {Barkhofen}}, \bibinfo {author}
  {\bibfnamefont {T.~J.}\ \bibnamefont {Bartley}},\ and\ \bibinfo {author}
  {\bibfnamefont {C.}~\bibnamefont {Silberhorn}},\ }\bibfield  {title}
  {\bibinfo {title} {{A high dynamic range optical detector for measuring
  single photons and bright light}},\ }\href
  {https://doi.org/10.1364/OE.27.000001} {\bibfield  {journal} {\bibinfo
  {journal} {Optics Express}\ }\textbf {\bibinfo {volume} {27}},\ \bibinfo
  {pages} {1} (\bibinfo {year} {2019})}\BibitemShut {NoStop}%
\bibitem [{\citenamefont {Lundeen}\ \emph {et~al.}(2009)\citenamefont
  {Lundeen}, \citenamefont {Feito}, \citenamefont {Coldenstrodt-Ronge},
  \citenamefont {Pregnell}, \citenamefont {Silberhorn}, \citenamefont {Ralph},
  \citenamefont {Eisert}, \citenamefont {Plenio},\ and\ \citenamefont
  {Walmsley}}]{lundeen2009tomography}%
  \BibitemOpen
  \bibfield  {author} {\bibinfo {author} {\bibfnamefont {J.~S.}\ \bibnamefont
  {Lundeen}}, \bibinfo {author} {\bibfnamefont {A.}~\bibnamefont {Feito}},
  \bibinfo {author} {\bibfnamefont {H.}~\bibnamefont {Coldenstrodt-Ronge}},
  \bibinfo {author} {\bibfnamefont {K.~L.}\ \bibnamefont {Pregnell}}, \bibinfo
  {author} {\bibfnamefont {C.}~\bibnamefont {Silberhorn}}, \bibinfo {author}
  {\bibfnamefont {T.~C.}\ \bibnamefont {Ralph}}, \bibinfo {author}
  {\bibfnamefont {J.}~\bibnamefont {Eisert}}, \bibinfo {author} {\bibfnamefont
  {M.~B.}\ \bibnamefont {Plenio}},\ and\ \bibinfo {author} {\bibfnamefont
  {I.~A.}\ \bibnamefont {Walmsley}},\ }\bibfield  {title} {\bibinfo {title}
  {{Tomography of quantum detectors}},\ }\href
  {https://doi.org/10.1038/nphys1133} {\bibfield  {journal} {\bibinfo
  {journal} {Nature Physics}\ }\textbf {\bibinfo {volume} {5}},\ \bibinfo
  {pages} {27} (\bibinfo {year} {2009})}\BibitemShut {NoStop}%
\bibitem [{\citenamefont {Nehra}\ \emph {et~al.}(2020)\citenamefont {Nehra},
  \citenamefont {Chang}, \citenamefont {Yu}, \citenamefont {Beling},\ and\
  \citenamefont {Pfister}}]{nehra2020photon}%
  \BibitemOpen
  \bibfield  {author} {\bibinfo {author} {\bibfnamefont {R.}~\bibnamefont
  {Nehra}}, \bibinfo {author} {\bibfnamefont {C.-H.}\ \bibnamefont {Chang}},
  \bibinfo {author} {\bibfnamefont {Q.}~\bibnamefont {Yu}}, \bibinfo {author}
  {\bibfnamefont {A.}~\bibnamefont {Beling}},\ and\ \bibinfo {author}
  {\bibfnamefont {O.}~\bibnamefont {Pfister}},\ }\bibfield  {title} {\bibinfo
  {title} {Photon-number-resolving segmented detectors based on single-photon
  avalanche-photodiodes},\ }\href {https://doi.org/10.1364/OE.380416}
  {\bibfield  {journal} {\bibinfo  {journal} {Optics Express}\ }\textbf
  {\bibinfo {volume} {28}},\ \bibinfo {pages} {3660} (\bibinfo {year}
  {2020})}\BibitemShut {NoStop}%
\bibitem [{\citenamefont {Miatto}\ \emph {et~al.}(2018)\citenamefont {Miatto},
  \citenamefont {Safari},\ and\ \citenamefont {Boyd}}]{miatto2018explicit}%
  \BibitemOpen
  \bibfield  {author} {\bibinfo {author} {\bibfnamefont {F.~M.}\ \bibnamefont
  {Miatto}}, \bibinfo {author} {\bibfnamefont {A.}~\bibnamefont {Safari}},\
  and\ \bibinfo {author} {\bibfnamefont {R.~W.}\ \bibnamefont {Boyd}},\
  }\bibfield  {title} {\bibinfo {title} {{Explicit formulas for photon number
  discrimination with on/off detectors}},\ }\href
  {https://doi.org/10.1364/AO.57.006750} {\bibfield  {journal} {\bibinfo
  {journal} {Applied Optics}\ }\textbf {\bibinfo {volume} {57}},\ \bibinfo
  {pages} {6750} (\bibinfo {year} {2018})}\BibitemShut {NoStop}%
\bibitem [{\citenamefont {Schapeler}\ \emph {et~al.}(2021)\citenamefont
  {Schapeler}, \citenamefont {Höpker},\ and\ \citenamefont
  {Bartley}}]{schapeler2021quantum}%
  \BibitemOpen
  \bibfield  {author} {\bibinfo {author} {\bibfnamefont {T.}~\bibnamefont
  {Schapeler}}, \bibinfo {author} {\bibfnamefont {J.~P.}\ \bibnamefont
  {Höpker}},\ and\ \bibinfo {author} {\bibfnamefont {T.~J.}\ \bibnamefont
  {Bartley}},\ }\bibfield  {title} {\bibinfo {title} {Quantum detector
  tomography of a high dynamic-range superconducting nanowire single-photon
  detector},\ }\href {https://doi.org/10.1088/1361-6668/abee9a} {\bibfield
  {journal} {\bibinfo  {journal} {Superconductor Science and Technology}\
  }\textbf {\bibinfo {volume} {34}},\ \bibinfo {pages} {64002} (\bibinfo {year}
  {2021})}\BibitemShut {NoStop}%
\bibitem [{\citenamefont {Feito}\ \emph {et~al.}(2009)\citenamefont {Feito},
  \citenamefont {Lundeen}, \citenamefont {Coldenstrodt-Ronge}, \citenamefont
  {Eisert}, \citenamefont {Plenio},\ and\ \citenamefont
  {Walmsley}}]{feito2009measuring}%
  \BibitemOpen
  \bibfield  {author} {\bibinfo {author} {\bibfnamefont {A.}~\bibnamefont
  {Feito}}, \bibinfo {author} {\bibfnamefont {J.~S.}\ \bibnamefont {Lundeen}},
  \bibinfo {author} {\bibfnamefont {H.}~\bibnamefont {Coldenstrodt-Ronge}},
  \bibinfo {author} {\bibfnamefont {J.}~\bibnamefont {Eisert}}, \bibinfo
  {author} {\bibfnamefont {M.~B.}\ \bibnamefont {Plenio}},\ and\ \bibinfo
  {author} {\bibfnamefont {I.~A.}\ \bibnamefont {Walmsley}},\ }\bibfield
  {title} {\bibinfo {title} {{Measuring measurement: theory and practice}},\
  }\href {https://doi.org/10.1088/1367-2630/11/9/093038} {\bibfield  {journal}
  {\bibinfo  {journal} {New Journal of Physics}\ }\textbf {\bibinfo {volume}
  {11}},\ \bibinfo {pages} {93038} (\bibinfo {year} {2009})}\BibitemShut
  {NoStop}%
\bibitem [{\citenamefont {Schapeler}\ \emph {et~al.}(2020)\citenamefont
  {Schapeler}, \citenamefont {H{\"{o}}pker},\ and\ \citenamefont
  {Bartley}}]{schapeler2020quantum}%
  \BibitemOpen
  \bibfield  {author} {\bibinfo {author} {\bibfnamefont {T.}~\bibnamefont
  {Schapeler}}, \bibinfo {author} {\bibfnamefont {J.~P.}\ \bibnamefont
  {H{\"{o}}pker}},\ and\ \bibinfo {author} {\bibfnamefont {T.~J.}\ \bibnamefont
  {Bartley}},\ }\bibfield  {title} {\bibinfo {title} {{Quantum detector
  tomography of a 2$\times$2 multi-pixel array of superconducting nanowire
  single photon detectors}},\ }\href {https://doi.org/10.1364/OE.404285}
  {\bibfield  {journal} {\bibinfo  {journal} {Optics Express}\ }\textbf
  {\bibinfo {volume} {28}},\ \bibinfo {pages} {33035} (\bibinfo {year}
  {2020})}\BibitemShut {NoStop}%
\bibitem [{\citenamefont {Schapeler}(2020)}]{schapeler2020cvxpycode}%
  \BibitemOpen
  \bibfield  {author} {\bibinfo {author} {\bibfnamefont {T.}~\bibnamefont
  {Schapeler}},\ }\href@noop {} {\bibinfo {title} {{Detector Tomography Python
  Code}}} (\bibinfo {year} {2020}),\ \bibinfo {note}
  {\url{https://physik.uni-paderborn.de/fileadmin/physik/Arbeitsgruppen/bartley/Downloads/Detector-Tomography-CVXPY-Python-Timon-Schapeler.zip}}\BibitemShut
  {NoStop}%
\bibitem [{\citenamefont {Diamond}\ and\ \citenamefont {Boyd}(2016)}]{cvxpy}%
  \BibitemOpen
  \bibfield  {author} {\bibinfo {author} {\bibfnamefont {S.}~\bibnamefont
  {Diamond}}\ and\ \bibinfo {author} {\bibfnamefont {S.}~\bibnamefont {Boyd}},\
  }\bibfield  {title} {\bibinfo {title} {{{CVXPY}: A {P}ython-Embedded Modeling
  Language for Convex Optimization}},\ }\href@noop {} {\bibfield  {journal}
  {\bibinfo  {journal} {Journal of Machine Learning Research}\ }\textbf
  {\bibinfo {volume} {17}},\ \bibinfo {pages} {1} (\bibinfo {year}
  {2016})}\BibitemShut {NoStop}%
\bibitem [{\citenamefont {Agrawal}\ \emph {et~al.}(2018)\citenamefont
  {Agrawal}, \citenamefont {Verschueren}, \citenamefont {Diamond},\ and\
  \citenamefont {Boyd}}]{cvxpy_rewriting}%
  \BibitemOpen
  \bibfield  {author} {\bibinfo {author} {\bibfnamefont {A.}~\bibnamefont
  {Agrawal}}, \bibinfo {author} {\bibfnamefont {R.}~\bibnamefont
  {Verschueren}}, \bibinfo {author} {\bibfnamefont {S.}~\bibnamefont
  {Diamond}},\ and\ \bibinfo {author} {\bibfnamefont {S.}~\bibnamefont
  {Boyd}},\ }\bibfield  {title} {\bibinfo {title} {{A rewriting system for
  convex optimization problems}},\ }\href
  {https://doi.org/10.1080/23307706.2017.1397554} {\bibfield  {journal}
  {\bibinfo  {journal} {Journal of Control and Decision}\ }\textbf {\bibinfo
  {volume} {5}},\ \bibinfo {pages} {42} (\bibinfo {year} {2018})}\BibitemShut
  {NoStop}%
\bibitem [{\citenamefont {Sperling}\ \emph {et~al.}(2012)\citenamefont
  {Sperling}, \citenamefont {Vogel},\ and\ \citenamefont
  {Agarwal}}]{sperling2012true}%
  \BibitemOpen
  \bibfield  {author} {\bibinfo {author} {\bibfnamefont {J.}~\bibnamefont
  {Sperling}}, \bibinfo {author} {\bibfnamefont {W.}~\bibnamefont {Vogel}},\
  and\ \bibinfo {author} {\bibfnamefont {G.~S.}\ \bibnamefont {Agarwal}},\
  }\bibfield  {title} {\bibinfo {title} {{True photocounting statistics of
  multiple on-off detectors}},\ }\href
  {https://doi.org/10.1103/PhysRevA.85.023820} {\bibfield  {journal} {\bibinfo
  {journal} {Physical Review A}\ }\textbf {\bibinfo {volume} {85}},\ \bibinfo
  {pages} {23820} (\bibinfo {year} {2012})}\BibitemShut {NoStop}%
\bibitem [{\citenamefont {J{\"{o}}nsson}\ and\ \citenamefont
  {Bj{\"{o}}rk}(2019)}]{jonsson2019evaluating}%
  \BibitemOpen
  \bibfield  {author} {\bibinfo {author} {\bibfnamefont {M.}~\bibnamefont
  {J{\"{o}}nsson}}\ and\ \bibinfo {author} {\bibfnamefont {G.}~\bibnamefont
  {Bj{\"{o}}rk}},\ }\bibfield  {title} {\bibinfo {title} {{Evaluating the
  performance of photon-number-resolving detectors}},\ }\href
  {https://doi.org/10.1103/PhysRevA.99.043822} {\bibfield  {journal} {\bibinfo
  {journal} {Physical Review A}\ }\textbf {\bibinfo {volume} {99}},\ \bibinfo
  {pages} {43822} (\bibinfo {year} {2019})}\BibitemShut {NoStop}%
\end{thebibliography}%

\end{document}